%% file: main.tex
\documentclass{article}
\usepackage{ijcai23}

\usepackage{times}
\usepackage{soul}
\usepackage{url}
\usepackage[hidelinks]{hyperref}
\usepackage[utf8]{inputenc}
\usepackage[small]{caption}
\usepackage{graphicx}
\usepackage{amsmath}
\usepackage{amsthm}
\usepackage{booktabs}
\usepackage[switch]{lineno}

\usepackage{booktabs} 

\usepackage[english]{babel}
\usepackage{moresize}
\usepackage{amsmath}

\usepackage{balance}
\usepackage{comment}
\usepackage{paralist}
\usepackage{multirow}

\usepackage{filecontents}

\usepackage[english]{babel}
\usepackage{mathrsfs}
\usepackage{graphicx}
\usepackage{amssymb}
\usepackage{url}
\usepackage{subfigure}
\usepackage{amsmath}
\usepackage{enumitem}
\usepackage{adjustbox}
\usepackage{amssymb}
\usepackage{times}
\usepackage{hyperref}

\usepackage{pgfplots}
\usetikzlibrary{pgfplots.dateplot}

\usepackage{filecontents}
\definecolor{tblue}{RGB}{31,119,180}
\definecolor{torange}{RGB}{255,127,14}
\definecolor{tgreen}{RGB}{44,160,44}
\definecolor{tred}{RGB}{214,39,40}
\definecolor{tpurple}{RGB}{148,103,189}

\usepackage{filecontents}

\usepackage{algorithm}
\usepackage{algorithmic}
\newcommand{\hide}[1]{} 

\newcommand{\ie}{\textit{i}.\textit{e}.}
\newcommand{\eg}{\textit{e}.\textit{g}.} 
\newcommand{\wrt}{\textit{w}.\textit{r}.\textit{t}}

\def\model{DSL}

\title{Denoised Self-Augmented Learning for Social Recommendation}

\author{Tianle Wang, Lianghao Xia, Chao Huang\thanks{Chao Huang is the Corresponding Author}\\
\affiliations
University of Hong Kong, Hong Kong\\
{louiswong.cs@connect.hku.hk, aka\_xia@foxmail.com \\ chaohuang75@gmail.com
}}

\begin{document}

\maketitle

\begin{abstract}
Social recommendation is gaining increasing attention in various online applications, including e-commerce and online streaming, where social information is leveraged to improve user-item interaction modeling. Recently, Self-Supervised Learning (SSL) has proven to be remarkably effective in addressing data sparsity through augmented learning tasks. Inspired by this, researchers have attempted to incorporate SSL into social recommendation by supplementing the primary supervised task with social-aware self-supervised signals. However, social information can be unavoidably noisy in characterizing user preferences due to the ubiquitous presence of interest-irrelevant social connections, such as colleagues or classmates who do not share many common interests. To address this challenge, we propose a novel social recommender called the \underline{D}enoised \underline{S}elf-Augmented \underline{L}earning paradigm (\model). Our model not only preserves helpful social relations to enhance user-item interaction modeling but also enables personalized cross-view knowledge transfer through adaptive semantic alignment in embedding space. Our experimental results on various recommendation benchmarks confirm the superiority of our \model\ over state-of-the-art methods. We release our model implementation at: \url{https://github.com/HKUDS/DSL}.

\end{abstract}

\input{intro}
\input{model}
\input{solution}
\input{eval}

\input{conclusion}


\clearpage
\bibliographystyle{named}
\bibliography{ijcai23}

\end{document}

%% file: intro.tex
\section{Introduction}
\label{sec:intro}


Social recommendation is a widely-used technique to improve the quality of recommender systems by incorporating social information into user preference learning~\cite{yu2021socially}. To accomplish this, various neural network techniques have been developed to encode social-aware user preferences for recommendation. Currently, the most advanced social recommendation methods are built using Graph Neural Networks (GNNs) for recursive message passing, which enables the capture of high-order correlations~\cite{fan2019graph,wu2019neural,song2019session}. In these architectures, user representations are refined by integrating information from both social and interaction neighbors.


While supervised GNN-enhanced models have achieved remarkable performance in social recommendation, they require a large amount of supervised labels to generate accurate user representations. In practical social recommendation scenarios, however, user-item interaction data is often very sparse~\cite{wei2022contrastive,chen2023heterogeneous}. This label sparsity severely limits the representation power of deep social recommenders and hinders their ability to reach their full potential. Recently, Self-Supervised Learning (SSL) has gained success due to its ability to avoid heavy reliance on observed label data in various domains, \eg, computer vision~\cite{he2020momentum}, natural language processing~\cite{liu2021simcls}, and graph representation learning~\cite{zhu2021graph}.

Motivated by the limitations of supervised GNN-enhanced models, recent attempts have adopted the self-supervised learning framework~\cite{yu2021socially}. These approach introduce an auxiliary learning task to supplement the supervised main task for data augmentation. For example, MHCN~\cite{yu2021self} uses a hypergraph-enhanced self-supervised learning framework to improve global relation learning in social recommender systems. Additionally, SMIN~\cite{long2021social} constructs metapath-guided node connections to explore the isomorphic transformation property of graph topology with augmented self-supervision signals.


Despite the decent performance of self-supervised learning, we argue that the SSL-based augmentation is severely hindered by noisy social relations when enhancing the representation learning of complex user preferences. While observed user-user social ties have the potential to capture social influence on user-item interaction behaviors, the model's performance can significantly degrade when trained on social-aware collaborative graphs with noisy social information. For instance, people may establish social connections with colleagues, classmates, or family members, but they may not share many common interests with each other~\cite{liu2019single,epasto2019single}. Therefore, these noisy social influences may not align with user preferences in real-life recommendation scenarios. In most existing solutions, information aggregated from noisy social neighbors may mislead graph message passing and self-supervised learning, resulting in sub-optimal recommendation performance.

To address the limitations mentioned earlier, we propose the Denoised Self-Augmented Learning (\model) paradigm for social recommender systems. Our approach leverages social information to better characterize user preferences with noise-resistant self-supervised learning, aimed at pursuing cross-view alignment. Firstly, we develop a dual-view graph neural network to encode latent representations over both user social and interaction graphs. Then, to mitigate the bias of social relations for recommendation, we design a denoising module to enhance the integrated social-aware self-supervised learning task. This module identifies unreliable user-wise connections with respect to their interaction patterns. Our \model\ is aware of the interaction commonality between users and can automate the social effect denoising process with adaptive user representation alignment. By doing so, the social-aware uniformity is well preserved in the learned user embeddings by alleviating the impact of noisy social information in our recommender.

Key contributions of this work are summarized as follows:

\begin{itemize}[leftmargin=*]


\item In this work, we investigate denoised self-augmented learning for social recommendation, effectively reducing the impact of noisy social relations on the representation of socially-aware collaborative signals.


\item We propose \model, which enables denoised cross-view alignment between the encoded embeddings from social and interaction views. The denoising module assigns reliability weights to useful social relations to encode user preference, endowing our \model\ with the capability of generating adaptive self-supervised signals.


\item We instantiate \model\ for social recommendation on three real-world datasets. Experimental results show that our method provides more accurate recommendations and superior performance in dealing with noisy and sparse data, compared with various state-of-the-art solutions.

\end{itemize}


%% file: model.tex
\section{Preliminaries and Related Work}
\label{sec:model}

\subsection{Social-aware Recommendation}
We denote the sets of users and items as $\mathcal{U}=\{u_1, ..., u_I\}$ and $\mathcal{V}=\{v_1, ..., v_J\}$, respectively, where $I$ and $J$ represent the number of users and items. The user-item interaction data is represented by an interaction graph $\mathcal{G}_{r}=\{\mathcal{U}, \mathcal{V}, \mathcal{E}_{r}\}$, where user-item connection edges are generated when user $u_i$ interacts with item $v_j$. To incorporate social context into the recommender system, we define a user social graph $\mathcal{G}_{s}=\{\mathcal{U}, \mathcal{E}_{s}\}$ to contain user-wise social connections. Social recommender systems aim to learn a model from both the user-item interaction graph $\mathcal{G}_{r}=\{\mathcal{U}, \mathcal{V}, \mathcal{E}_{r}\}$ and the user-user social graph $\mathcal{G}_{s}=\{\mathcal{U}, \mathcal{E}_{s}\}$, encoding user interests to make accurate recommendations.


In recent years, several neural network techniques have been proposed to solve the social recommendation problem. For instance, attention mechanisms have been used to differentiate social influence among users with learned attentional weights, as seen in EATNN~\cite{chen2019efficient} and SAMN~\cite{chen2019social}. Many GNN-based social recommender systems have been developed to jointly model user-user and user-item graphs via message passing, leveraging the effectiveness of high-order relation encoding with graph neural networks~\cite{wang2019neural}. Examples include GraphRec~\cite{fan2019graph}, DiffNet~\cite{wu2019neural}, and FeSoG~\cite{liu2022federated}. Some recent attempts have leveraged self-supervised learning to enhance social recommendation with auxiliary self-supervision signals, such as MHCN~\cite{yu2021self} and SMIN~\cite{long2021social}. However, their representation learning abilities are limited by social relation noise, which leads to biased models.


\subsection{Self-Supervised Recommender Systems}
Self-supervised learning has recently gained attention in various recommendation tasks. Supervised contrastive loss has been shown to benefit graph collaborative filtering with effective data augmentation~\cite{wu2021self,cailightgcl}. For sequential recommender systems, self-supervised pre-training~\cite{zhou2020s3} and imitation~\cite{yuan2021improving} have been introduced to enhance sequence modeling. Researchers have also brought the benefits of self-supervised learning to multi-interest/multi-behavior recommender systems, as seen in Re4~\cite{zhang2022re4} and CML~\cite{weicmlcontrastive}. Our method advances this research by proposing a novel unbiased self-supervised learning paradigm to denoise social relation modeling in social recommender systems.

%% file: solution.tex
\section{Methodology}
\label{sec:solution}

\begin{figure*}[htbp]
    \centering
    \includegraphics[width=\textwidth,height=0.35\textwidth]{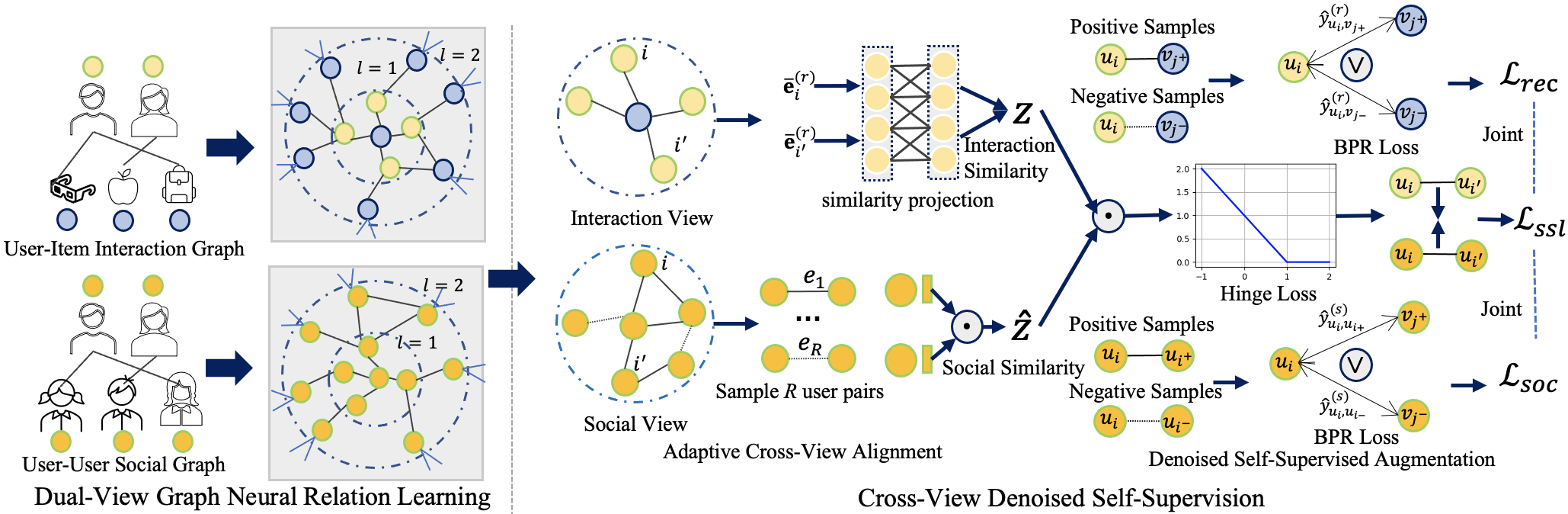}
    \caption{Overall framework of the proposed denoised self-augmented learning (\model) model.}
    \label{fig:model}
\end{figure*}

In this section, we present our \model\ model with technical details. The model architecture is illustrated in Figure~\ref{fig:model}.


\subsection{Dual-View Graph Neural Relation Learning}
With the initialized id-corresponding embeddings, our \model\ first employs a dual-view graph neural network to capture high-order collaborative relations for both user-item interactions and user-user social ties. Inspired by the effectiveness of lightweight GCN-enhanced collaborative filtering paradigms~\cite{he2020lightgcn,chen2020revisiting}, \model\ is configured with a simplified graph neural network, which is:
\begin{align}
    \textbf{E}^{(l)}_r = (\mathcal{L}_r+\textbf{I})\cdot \textbf{E}^{(l-1)}_r
\end{align}
\noindent The above equation shows the iterative information propagation scheme of our GCN over the user-item interaction graph. Here, $\textbf{E}^{(l)}_r,\textbf{E}_r^{(l-1)}\in\mathbb{R}^{(I+J)\times d}$ denote the embeddings of users and items after $l$ iterations of user-item relation modeling. $\textbf{E}^{(0)}_r$ is initialized by stacking the initial user embedding matrix $\textbf{E}_u$ and the item embedding matrix $\textbf{E}_v$. $\textbf{I}\in\mathbb{R}^{(I+J)\times (I+J)}$ denotes the identity matrix for enabling self-loop. $\mathcal{L}_r\in\mathbb{R}^{(I+J)\times (I+J)}$ denotes the Laplacian matrix of the user-item interaction graph~\cite{wang2019neural}.
\begin{align}
    \mathcal{L}_r=\textbf{D}_r^{-\frac{1}{2}}\textbf{A}_r\textbf{D}_r^{-\frac{1}{2}},~~~~ \textbf{A}_r=\left[\begin{array}{cc}\textbf{0} & \textbf{R} \\\textbf{R}^{\top} &\textbf{0}\end{array}\right]
\end{align}
\noindent $\textbf{R}\in\mathbb{R}^{I\times J}$ denotes the user-item interaction matrix, and $\textbf{0}$ refers to all-zero matrices. The bidirectional adjacent matrix $\textbf{A}_r$ of the user-item interaction view is multiplied by its corresponding diagonal degree matrix $\textbf{D}_r$ for normalization.

To encode user-wise social relations in our recommender, we also apply the lightweight GCN to the user social graph $\mathcal{G}_{s}$. Specifically, our social view GNN takes the initial users' id-corresponding embeddings as input by setting $\textbf{E}^{(0)}_s=\textbf{E}_u$. The user embeddings are generated by cross-layer passing:
\begin{align}
    \textbf{E}^{(l)}_s=(\mathcal{L}_s+\textbf{I})\cdot\textbf{E}_s^{(l-1)},~~~~\mathcal{L}_s=\textbf{D}_s^{-\frac{1}{2}}\textbf{S}~\textbf{D}_s^{-\frac{1}{2}}
\end{align}
\noindent Here, $\textbf{S}\in\mathbb{R}^{I\times I}$ encodes the user-wise social relations, and $\textbf{D}_s, \mathcal{L}_s\in\mathbb{R}^{I\times I}$ denote the corresponding diagonal degree matrix and the normalized Laplacian matrix for the social view. $\textbf{E}_s^{(l)},\textbf{E}_s^{(l-1)}\in\mathbb{R}^{I\times d}$ are the users' social embeddings in the $l$-th and $(l-1)$-th graph neural iteration, respectively. \\ \vspace{-0.1in}


\paragraph{Embedding Aggregation.} To aggregate the embeddings encoded from different orders in $\mathcal{G}_{r}$ and $\mathcal{G}_{s}$, \model\ adopts mean-pooling operators for both the interaction and social views.
\begin{align}
    \bar{\textbf{E}}_r = \sum_{l=0}^L\textbf{E}_r^{(l)},~~~~\bar{\textbf{E}}_s = \sum_{l=0}^L\textbf{E}_s^{(l)}
\end{align}
\noindent Here, $L$ is the maximum number of graph iterations. With our dual-view GNN, we encode view-specific relations for user interaction and social influence in our model.


\subsection{Cross-View Denoised Self-Supervision}
In our recommender, the learned user-item relations and user-wise dependencies are complementary to each other. To integrate both relational contextual signals, we design a cross-view denoised self-supervised learning paradigm that can alleviate the noisy effects of transferring social knowledge into user-item interaction modeling. In real-life scenarios, passively-built social relations, such as colleagues or classmates, may not bring much influence to user interaction preference due to their diverse shopping tastes. Blindly relying on such irrelevant social ties to infer users' interests could damage the performance of social recommendation models. To address this issue, we filter out the noisy social influence between dissimilar users with respect to their interaction preference for unbiased self-supervision. \\ \vspace{-0.1in}



\paragraph{Adaptive Cross-View Alignment.} 
In our \model, we incorporate the cross-view denoising task to supplement the main learning task with auxiliary self-supervision signals. The learned user interaction patterns guide the social relation denoising module to filter out misleading embedding propagation based on observed social connections. Specifically, the interaction similarity $z_{i,i'}$ between the user pair ($i$, $i'$) is generated by $z_{i,i'} = [\bar{\textbf{e}}^{(r)}_i;\bar{\textbf{e}}^{(r)}_{i'}]$, given the user embeddings ($\bar{\textbf{e}}^{(r)}_i$, $\bar{\textbf{e}}^{(r)}_{i'}$) learned from our interaction GNN. Similarly, user social similarity $\hat{z}_{i,i'}$ can be obtained by $\hat{z}_{i,i'}=[\bar{\textbf{e}}_i^{(s)};\bar{\textbf{e}}_{i'}^{(s)}]$, based on user representations ($\bar{\textbf{e}}^{(s)}_i$, $\bar{\textbf{e}}^{(s)}_{i'}$) encoded from our social GNN. To alleviate the semantic gap between interaction view and social view, we design a learnable similarity projection function to map interaction semantics into a latent embedding space for cross-view alignment, as follows:
\begin{align}
z_{i,i'} = \text{sigm}(\textbf{d}^\top\cdot\sigma(\textbf{T}\cdot[\bar{\textbf{e}}^{(r)}_i;\bar{\textbf{e}}^{(r)}_{i'}]+\bar{\textbf{e}}^{(r)}_i+\bar{\textbf{e}}^{(r)}_{i'}+\textbf{c}))
\end{align}
\noindent where $\text{sigm}(\cdot)$ and $\sigma(\cdot)$ denote the sigmoid and LeakyReLU activation functions, respectively. Our designed parameterized projection function consists of $\textbf{d}\in\mathbb{R}^d, \textbf{T}\in\mathbb{R}^{d\times 2d}, \textbf{c}\in\mathbb{R}^d$ as learnable parameters, enabling adaptive alignment between social and interaction views. \\\vspace{-0.1in}

\paragraph{Denoised Self-Supervised Augmentation.} 
To incorporate denoised social influence to improve recommendation quality, we design a self-supervised learning task for cross-view alignment with augmented embedding regularization. Specifically, the cross-view alignment loss function is:
\begin{align}
    L_{ssl} = \sum_{(u_i,u_{i'})} \max (0, 1-z_{i,i'}\hat{z}_{i,i'})
\end{align}
\noindent The user pair ($u_i, u_{i'}$) is individually sampled from user set $\mathcal{U}$. With the above self-supervised learning objective, the integrated user relation prediction task will be guided based on the self-supervised signals for social influence denoising. Dy doing so, the noisy social connections between users with dissimilar preference, which contradicts with the target recommendation task will result in distinguishable user representations for recommendation enhancement.


\subsection{Multi-Task Model Optimization}
The learning process of our \model\ involves multi-task training for model optimization. The augmented self-supervised learning tasks are integrated with the main recommendation optimized loss to model denoised social-aware user preferences. Given the encoded user and item embeddings, we predict user-item ($\hat{y}_{u_i,v_j}^{(r)}$) and user-user ($\hat{y}_{u_i, u_{i'}}^{(s)}$) relations as:
\begin{align}
    \hat{y}_{u_i,v_j}^{(r)} = \bar{\textbf{e}}_i^{(r)\top} \bar{\textbf{e}}_j^{(r)};~~~~
    \hat{y}_{u_i, u_{i'}}^{(s)} = \bar{\textbf{e}}_i^{(s)\top} \bar{\textbf{e}}_{i'}^{(s)}
\end{align}
\noindent where $\hat{y}_{u_i,v_j}^{(r)} \in \mathbb{R}$ represents the likelihood of user $u_i$ interacting with item $v_j$ from the interaction view, while $\hat{y}_{u_i, u_{i'}}^{(s)}$ indicates the probability of $u_i$ and $u_{i'}$ being socially connected. Given these definitions, we minimize the following BPR loss functions~\cite{rendle2009bpr} for optimization:
\begin{align}
    &{L}_{rec} = \sum_{(u_i,v_{j^+}, v_{j^-})} - \ln \text{sigm}(\hat{y}_{u_i,v_{j^+}}^{(r)} - \hat{y}_{u_i,v_{j^-}}^{(r)})\nonumber\\
    &{L}_{soc} = \sum_{(u_i,u_{i^+}, u_{i^-})} - \ln \text{sigm}(\hat{y}_{u_i,u_{i^+}}^{(s)} - \hat{y}_{u_i,u_{i^-}}^{(s)})
\end{align}
\noindent where $v_{j^+}$ and $v_{j^-}$ denote the sampled positive and negative item for user $u_i$. $u_{i^+}$ and $u_{i^-}$ are sampled from $u_i$'s socially-connected and unconnected users, respectively. By integrating self-supervised learning objectives with weight parameter $\lambda_1, \lambda_2, \lambda_3$, the joint optimized loss is given as:
\begin{align}
    L = L_{rec} + \lambda_1 L_{soc} + \lambda_2 L_{ssl} + \lambda_3 (\|\textbf{E}_u\|_F^2 + \|\textbf{E}_v\|_F^2)
\end{align}

\subsection{In-Depth Analysis of \model}
In this section, our aim is to answer the question: \emph{How does our model enable adaptive and efficient self-supervised learning?} We provide analysis to further understand our model.\\\vspace{-0.12in}

\paragraph{Adaptive Self-Supervised Learning.}
In most existing contrastive learning (CL) approaches, auxiliary SSL signals are generated to address the issue of sparse supervision labels. Following the mutual information maximization (Infomax) principle, these approaches maximize the agreement between positive samples while pushing negative pairs away in the embedding space, as shown below with derived gradients.
\begin{align}
    \frac{\partial L_{cl}}{\partial \bar{\textbf{e}}_i} = -\bar{\textbf{e}}_{i^+} + \sum_{u_{i^-}} \bar{\textbf{e}}_{i^-} \frac{\exp\bar{\textbf{e}}_i^\top \bar{\textbf{e}}_{i^-}  }{\sum_{u_{i^-}} \exp \bar{\textbf{e}}_i^\top\bar{\textbf{e}}_{i^-}}
\end{align}
\noindent The first term in the equation maximizes the similarity between positive pairs ($\bar{\textbf{e}}_i$ and $\bar{\textbf{e}}_{i^+}$) with the same strength. Negative pairs ($\bar{\textbf{e}}_i$ and $\bar{\textbf{e}}_{i^-}$) with higher similarity are pushed away with greater strength, while negative pairs with lower similarity are pushed away with lesser strength. For simplicity, we have omitted the vector normalization and the temperature coefficient in the above-presented InfoNCE loss.

In comparison, our cross-view denoising SSL schema aims to maximize the similarity between sampled user pairs $(\bar{\textbf{e}}_i, \bar{\textbf{e}}_{i'})$ adaptively, based on the labels $z_{i,i'}$. The non-zero gradients of our denoising SSL over $\bar{\textbf{e}}_i$ are shown below:
\begin{align}
    \frac{\partial L_{ssl}}{\partial \bar{\textbf{e}}_i} = \sum_{u_{i'}} - z_{i,i'}\bar{\textbf{e}}_{i'}
\end{align}
\noindent The learnable $z_{i,i'}$ reflects the common preference between user $u_i$ and $u_{i'}$, which adaptively controls the strength of our self-supervised regularization. This enables us to filter out noisy signals in the observed social connections, and supercharge our SSL paradigm with adaptive data augmentation by transferring knowledge across different semantic views.
\\\vspace{-0.1in}

\paragraph{Efficient SSL.}
Our \model\ model adopts a lightweight graph convolutional network (GCN) as the graph relation encoder, with a complexity of $\mathcal{O}((|\mathcal{E}_r|+|\mathcal{E}_s|)\times d)$. The cross-view denoised self-supervised learning conducts pairwise node-wise relationships, which takes $\mathcal{O}(B\times d)$ time complexity, where $B$ represents the batch size. In contrast, most existing vanilla InfoNCE-based contrastive learning methods calculate relations between a batch of nodes and all other nodes, resulting in an operation complexity of $\mathcal{O}(B\times I\times d)$. 

%% file: eval.tex
\section{Evaluation}
\label{sec:eval}

\begin{table}[t]
\small
\centering
\begin{tabular}{c|c|c|c}
\hline
Data               & Ciao     & Epinions & Yelp      \\ \hline
\# Users            & 6,672    & 11,111   & 161,305   \\
\# Items            & 98,875   & 190,774  & 114,852   \\
\# Interactions     & 198,181  & 247,591  & 1,118,645 \\
Interaction Density & 0.0300\% & 0.0117\% & 0.0060\%  \\
\# Social Ties      & 109,503  & 203,989  & 2,142,242 \\ \hline
\end{tabular}
\caption{Statistical information of evaluated datasets.}
\label{tab:statistic}
\end{table}

We conduct extensive experiments to evaluate the effectiveness of our \model\ by answering the following research questions: \textbf{RQ1}: Does \model\ outperform state-of-the-art recommender systems? \textbf{RQ2}: How do different components affect the performance of \model? \textbf{RQ3}: Is \model\ robust enough to handle noisy and sparse data in social recommendation? \textbf{RQ4}: How efficient is \model\ compared to alternative methods?

\subsection{Experimental Settings}

\paragraph{Dataset.} 
We conduct experiments on three benchmark datasets collected from the Ciao, Epinions, and Yelp online platforms, where social connections can be established among users in addition to their observed implicit feedback (\eg, rating, click) over different items. Table~\ref{tab:statistic} lists the detailed statistical information of the experimented datasets. \\\vspace{-0.1in}


\paragraph{Metrics.} We use Hit Ratio (HR)@N and Normalized Discounted Cumulative Gain (NDCG)@N as evaluation metrics, where $N$ is set to 10 by default. We adopt a leave-one-out strategy, following similar settings as in~\cite{long2021social}.

\begin{table*}[!h]\small
\centering
\begin{tabular}{c|c|c|c|c|c|c|c|c|c|c|c|c}
\hline
Dataset                   & Metrics & PMF    & TrustMF & DiffNet & DGRec  & EATNN  & NGCF+  & MHCN   & KCGN    & SMIN   & \model\ & \%Imp            \\ \hline
\multirow{2}{*}{Ciao}     & HR      & 0.4223 & 0.4492  & 0.5544  & 0.4658 & 0.4255 & 0.5629 & 0.5950 & 0.5785 & 0.5852 & \textbf{0.6374} & 26.0\\ \cline{2-13} 
                          & NDCG    & 0.2464 & 0.2520  & 0.3167  & 0.2401 & 0.2525 & 0.3429 & 0.3805 & 0.3552 & 0.3687 & \textbf{0.4065} & 37.2 \\ \hline
\multirow{2}{*}{Epinions} & HR      & 0.1686 & 0.1769  & 0.2182  & 0.2055 & 0.1576 & 0.2969 & 0.3507 & 0.3122 & 0.3159 & \textbf{0.3983} & 76.9 \\ \cline{2-13} 
                          & NDCG    & 0.0968 & 0.0842  & 0.1162  & 0.0908 & 0.0794 & 0.1582 & 0.1926 & 0.1721 & 0.1867 & \textbf{0.2290} & 96.2 \\ \hline
\multirow{2}{*}{Yelp}     & HR      & 0.7554 & 0.7791  & 0.8031  & 0.7950 & 0.8031 & 0.8265 & 0.8571 & 0.8484 & 0.8478 & \textbf{0.8923} & 10.1 \\ \cline{2-13} 
                          & NDCG    & 0.5165 & 0.5424  & 0.5670  & 0.5593 & 0.5560 & 0.5854 & 0.6310 & 0.6028 & 0.5993 & \textbf{0.6599} & 15.5 \\ \hline
\end{tabular}
\vspace{-0.05in}
\caption{Recommendation performance of different methods. \%Imp denotes relative improvements over all baselines on average.}
\label{tab:baseline}
\end{table*}

\subsection{Baseline Methods}
We evaluate the performance of \model\ by comparing it with 10 baselines from different research lines for comprehensive evaluation, including: i) MF-based recommendation approaches (\ie, PMF, TrustMF); ii) attentional social recommenders (\ie, EATNN); iii) GNN-enhanced social recommendation methods (\ie, DiffNet, DGRec, NGCF+); and iv) self-supervised social recommendation models (\ie, MHCN, KCGN, SMIN, DcRec). Details are provided as follows:
\begin{itemize}[leftmargin=*]

\item \textbf{PMF}~\cite{mnih2007probabilistic}: is a probabilistic approach that uses matrix factorization technique to factorize users and items into latent vectors for representations.

\item \textbf{TrustMF}~\cite{yang2016social}: This method incorporates trust relations between users into matrix factorization as social information to improve recommendation performance.

\item \textbf{EATNN}~\cite{chen2019efficient}: It is an adaptive transfer learning model built upon attention mechanisms to aggregate information from both user interactions and social ties.

\item \textbf{DiffNet}~\cite{wu2019neural}: This is a deep influence propagation architecture to recursively update users' embeddings with social influence diffusion components.

\item \textbf{DGRec}~\cite{song2019session}: This social recommender leverages a graph attention network to jointly model the dynamic behavioral patterns of users and social influence.

\item \textbf{NGCF+}~\cite{wang2019neural}: This GNN-enhanced collaborative filtering approach performs message passing over a social-aware user-item relation graph.

\item \textbf{MHCN}~\cite{yu2021self}: This model proposes a multi-channel hypergraph convolutional network to enhance social recommendation by considering high-order relations.

\item \textbf{KCGN}~\cite{huang2021knowledge}: It improves social recommendation by integrating item inter-dependent knowledge with social influence through a multi-task learning framework.

\item \textbf{SMIN}~\cite{long2021social}: This model incorporates a metapath-guided heterogeneous graph learning task into social recommendation, utilizing self-supervised signals based on mutual information maximization.


\end{itemize}

\paragraph{Implementation Details.} We implement our \model\ using PyTorch and optimize parameter inference with Adam. During training, we use a learning rate range of $[5e^{-4}, 1e^{-3}, 5e^{-3}]$ and a decay ratio of $0.96$ per epoch. The batch size is selected from $[1024, 2048, 4096, 8192]$ and the hidden dimensionality is tuned from $[64, 128, 256, 512]$. We search for the optimal number of information propagation layers in our graph neural architecture from $[1, 2, 3, 4]$. The regularization weights $\lambda_1$ and $\lambda_2$ are selected from $[1e^{-3}, 1e^{-2}, 1e^{-1}, 1e^{0}, 1e^{1}]$ and $[1e^{-6}, 1e^{-5}, 1e^{-4}, 1e^{-3}]$, respectively. The weight for weight-decay regularization $\lambda_3$ is tuned from $[1e^{-7}, 1e^{-6}, 1e^{-5}, 1e^{-4}]$.

\subsection{Overall Performance Comparison (RQ1)}
Table~\ref{tab:baseline} and Table~\ref{tab:topn} demonstrate that our \model\ framework consistently outperforms all baselines on various datasets, providing evidence of its effectiveness. Based on our results, we make the following observations.

\begin{itemize}[leftmargin=*]

\item Our results demonstrate that \model\ achieves encouraging improvements on datasets with diverse characteristics, such as varying interaction densities. Specifically, on the Ciao and Epinions datasets, \model\ achieves an average improvement of 26.0\% and 76.0\% over baselines, respectively. This validates the importance of addressing noise issues in social information incorporation, which can effectively debias user representations and boost recommendation performance.

\item Methods incorporating self-supervised augmentation consistently outperform other baselines, highlighting the importance of exploring self-supervision signals from unlabeled data to alleviate sparsity issues in social recommendation. Our \model\ outperforms other methods, suggesting that denoising social relation modeling in socially-aware recommender systems can benefit the design of more helpful self-supervision information. In MHCN and SMIN, mutual information is maximized to reach agreement between socially-connected users under a self-supervised learning framework. However, blindly generating augmented self-supervision labels from noisy social connections can align embeddings of connected users, diluting their true preferences. In contrast, our \model\ can mitigate the effects of false positives for socially-dependent users.

\item GNN-enhanced social recommenders, \eg, DiffNet and DGRec, outperform vanilla attentive methods like EATNN, highlighting the effectiveness of modeling high-order connectivity in social-aware collaborative relationships. This observation aligns with the conclusion that incorporating high-hop information fusion is beneficial for embedding learning in CF signals. However, aggregating irrelevant social information via GNNs can lead to unwanted embedding propagation and weaken model representation ability.

\end{itemize}

\begin{table}[t]
\footnotesize
\centering
\begin{tabular}{c|cc|cc|cc}
\hline
Data & \multicolumn{2}{c|}{Ciao}                              & \multicolumn{2}{c|}{Epinions}                          & \multicolumn{2}{c}{Yelp}                               \\ \hline
Metrics  & \multicolumn{1}{c|}{\scriptsize{HR}}              & \scriptsize{NDCG}            & \multicolumn{1}{c|}{\scriptsize{HR}}             & \scriptsize{NDCG}            & \multicolumn{1}{c|}{\scriptsize{HR}}              & \scriptsize{NDCG}            \\ \hline
DSL-d    & \multicolumn{1}{c|}{0.615}          & 0.399          & \multicolumn{1}{c|}{0.354}          & 0.207          & \multicolumn{1}{c|}{0.887}          & 0.658          \\
DSL-s    & \multicolumn{1}{c|}{0.594}          & 0.374          & \multicolumn{1}{c|}{0.327}          & 0.169          & \multicolumn{1}{c|}{0.839}          & 0.621          \\
DSL-c    & \multicolumn{1}{c|}{0.603}          & 0.388          & \multicolumn{1}{c|}{0.336}          & 0.199          & \multicolumn{1}{c|}{0.889}          & \textbf{0.662} \\ \hline
DSL      & \multicolumn{1}{c|}{\textbf{0.637}} & \textbf{0.406} & \multicolumn{1}{c|}{\textbf{0.398}} & \textbf{0.229} & \multicolumn{1}{c|}{\textbf{0.892}} & 0.659         \\ \hline
\end{tabular}
\caption{Component-wise ablation study of \model.}
\label{tab:ablation}
\end{table}

\begin{table*}[h]\small
\centering
\begin{tabular}{c|c|c|c|c|c|c|c|c|c|c|c}
\hline
Metrics & PMF    & TrustMF & DiffNet & DGRec  & EATNN  & NGCF+  & MHCN   & KCGN   & SMIN   & \model\ &  \%Imp           \\ \hline
HR@5    & 0.3032 & 0.3133  & 0.3963  & 0.3035 & 0.3259 & 0.4263 & 0.4762 & 0.4361 & 0.4565 & \textbf{0.5007} & 35.2 \\ \hline
NDCG@5  & 0.2071 & 0.2073  & 0.2650  & 0.1872 & 0.2342 & 0.3020 & 0.3479 & 0.3094 & 0.3298 & \textbf{0.3626} & 43.0 \\ \hline
HR@10   & 0.4223 & 0.4492  & 0.5544  & 0.4658 & 0.4255 & 0.5629 & 0.5950 & 0.5785 & 0.5852 & \textbf{0.6374} & 26.0 \\ \hline
NDCG@10 & 0.2464 & 0.2520  & 0.3167  & 0.2401 & 0.2525 & 0.3429 & 0.3805 & 0.3552 & 0.3687 & \textbf{0.4065} & 37.2 \\ \hline
HR@20   & 0.5565 & 0.6133  & 0.6973  & 0.6193 & 0.5309 & 0.7032 & 0.7418 & 0.7191 & 0.7000 & \textbf{0.7683} & 18.6 \\ \hline
NDCG@20 & 0.2799 & 0.3020  & 0.3514  & 0.2746 & 0.2838 & 0.3675 & 0.4241 & 0.3844 & 0.3780 & \textbf{0.4353} & 31.6 \\ \hline
\end{tabular}
\caption{Ranking performance on Ciao dataset with varying Top-N values in terms of HR@N and NDCG@N}
\label{tab:topn}
\end{table*}

\subsection{Impact Study of Different Components (RQ2)}
In this section, we examine the effects of component-wise ablation and how hyperparameters influence performance. \\\vspace{-0.1in}

\paragraph{Model Ablation Study.}
To investigate the essential role of our denoised self-supervised learning paradigm in improving performance, we perform an ablation study of key model components. Specifically, we compare our \model\ with the following ablated variants: (1) ``\model-d'': disabling cross-view denoised self-supervised learning for mitigating the noisy effects of social-aware collaborative filtering, (2) ``\model-s'': removing social-aware self-supervised augmentation and directly incorporating social user embeddings into user-item interaction prediction, and (3) ``\model-c'': replacing denoised cross-view alignment with contrastive learning to reach agreement for integrating social and interaction views. Results are reported in Table~\ref{tab:ablation}.
From our results, we observe that our \model\ outperforms other variants in most evaluation cases. Based on this, we draw the following key conclusions:
\begin{itemize}[leftmargin=*]

\item Comparing \model\ with ``\model-d'', the significant performance improvement suggests that social-aware collaborative relation learning is affected by the noise problem when directly incorporating social information.

\item The recommendation performance further drops in variant ``\model-s'' without the auxiliary learning task of social-aware relation prediction, indicating that incorporating self-supervised signals from user-wise social influence is helpful for enhancing collaborative relational learning.

\item The contrastive learning used in variant ``\model-c'' attempts to align the social and interaction views, but the irrelevant social connections can mislead the contrastive self-supervision for data augmentation. This observation supports our assumption that social information is inherently noisy for characterizing user preferences.

\end{itemize}

\begin{figure}[t]
\centering
\subfigure{
    \includegraphics[width=0.15\textwidth]{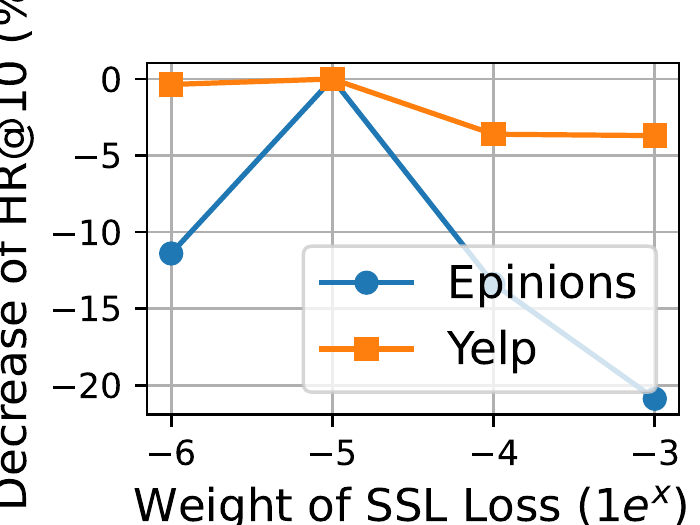}
    \includegraphics[width=0.15\textwidth]{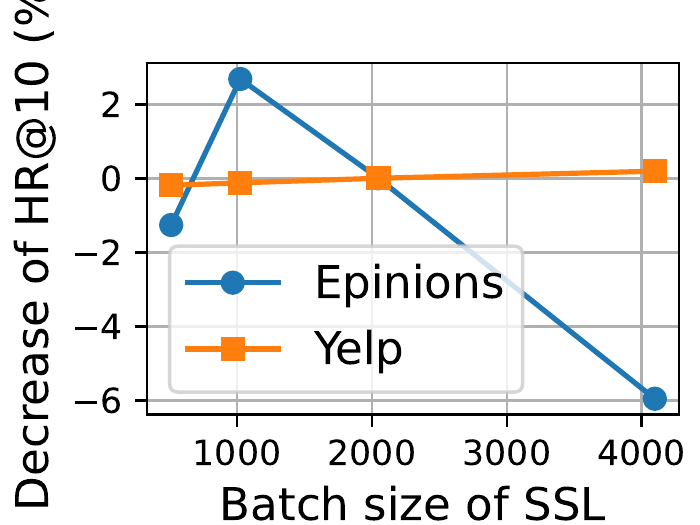}
    \includegraphics[width=0.15\textwidth]{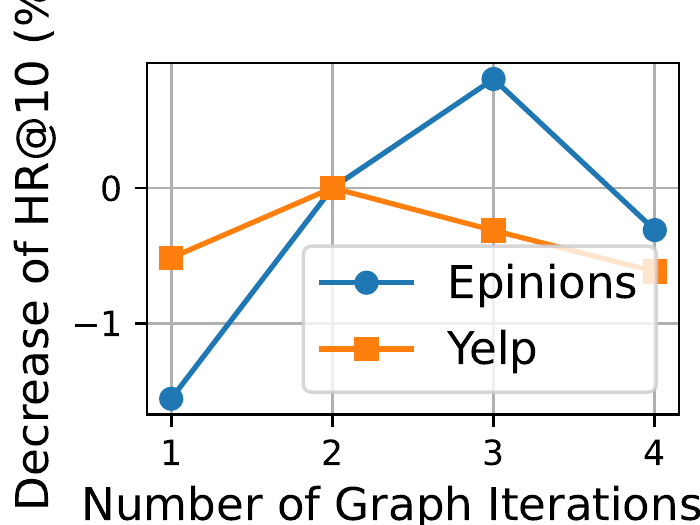}
}
\vspace{-0.1in}
\caption{We conduct a hyperparameter study of the \model\ with respect to i) SSL loss weight for regularization, ii) batch size for training, and iii) \# of propagation layers for message passing.}
\label{fig:parameter}
\end{figure}

\paragraph{Parameter Effect Study.}
Exploring the influence of key hyperparameters on \model's performance, including SSL loss weight, batch size, and the number of graph propagation layers, would be interesting. The results are shown in Figure~\ref{fig:parameter}, where the y-axis represents the performance variation ratio compared to the default parameter settings.
\begin{itemize}[leftmargin=*]

\item \textbf{Effect of SSL regularization weight}. The SSL loss weight controls the regularization strength of self-supervision signals. It is clear that a proper weight of SSL loss regularization is beneficial for improving model learning on highly-skewed distributed data. However, the SSL regularization does not always improve model representation. As the SSL loss weight increases, the performance worsens. This is because the model gradient learning is biased towards the strongly regularized SSL signals, which has negative effects on the main optimized objective for recommendation.

\item \textbf{Effect of batch size}. The best model performance is achieved with a batch size of around 2048. The observed performance differences between Epinions and Yelp stem from their diverse social data densities. Epinions data is more sensitive to batch size due to its sparse social connections. Results on Epinions data indicate that a larger batch size helps alleviate the over-fitting issue during model training. However, worse performance is observed with further increasing batch size, possibly due to local optima.

\item \textbf{Effect of propagation layers \#}. In GNNs, the number of propagation layers balances the trade-off between informativeness and over-smoothing. When tuning $L$ from 1 to 4, other parameters are kept at their default settings. A deeper graph neural network is effective in modeling high-order connectivity through cross-layer message passing to generate informative user and item representations. However, stacking too many propagation layers reduces the model capacity by making many different users identical. The resulting over-smoothed embeddings cannot preserve user uniformity with discriminative representations.

\end{itemize}

\begin{figure*}[t!]
\centering
    \includegraphics[width=0.19\textwidth]{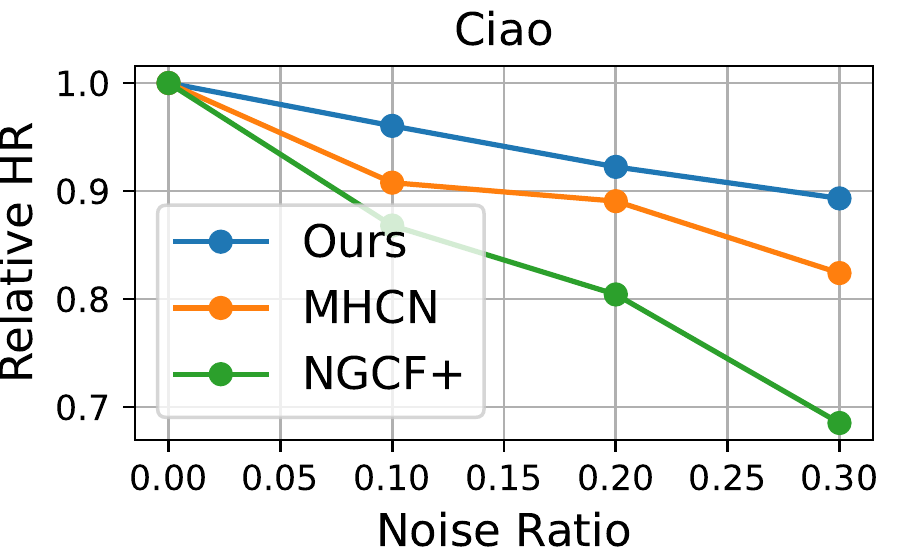}
    \includegraphics[width=0.19\textwidth]{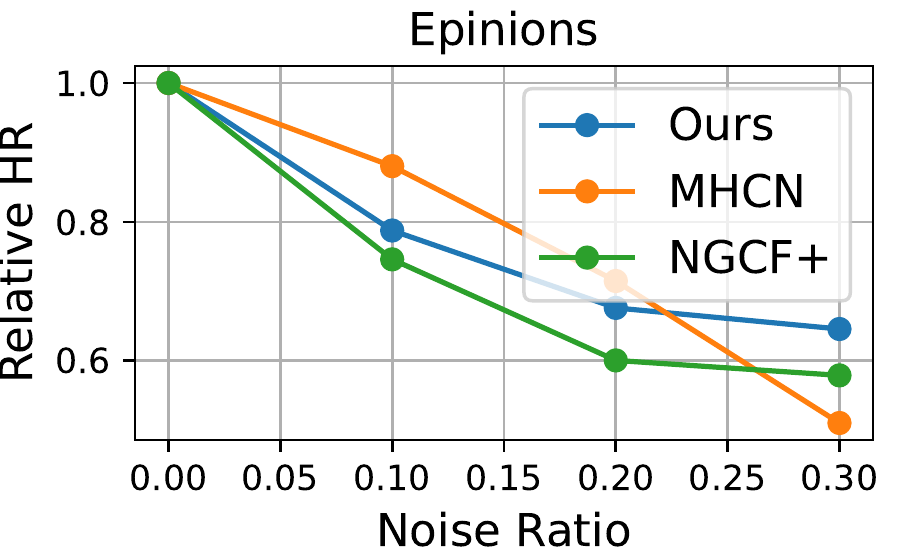}
    \includegraphics[width=0.19\textwidth]{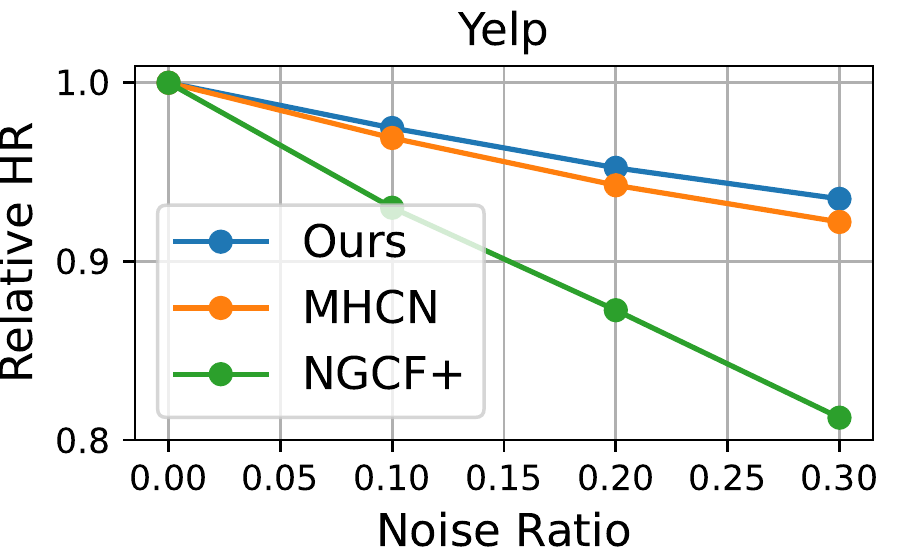}
    \includegraphics[width=0.19\textwidth]{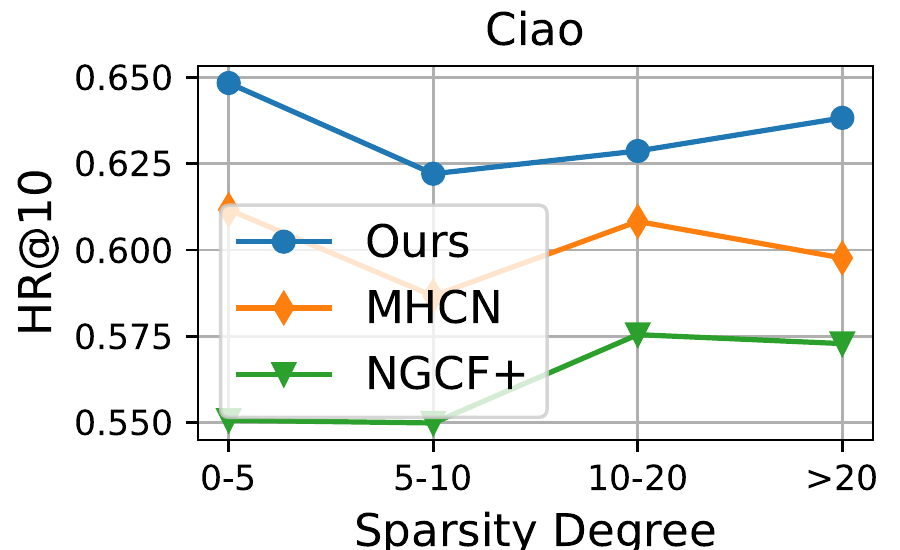}
    \includegraphics[width=0.19\textwidth]{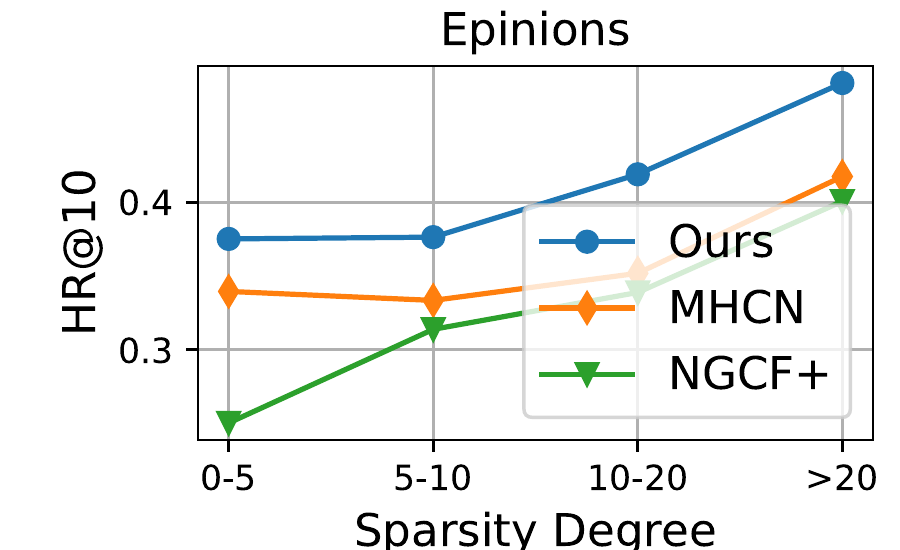}
    \includegraphics[width=0.19\textwidth]{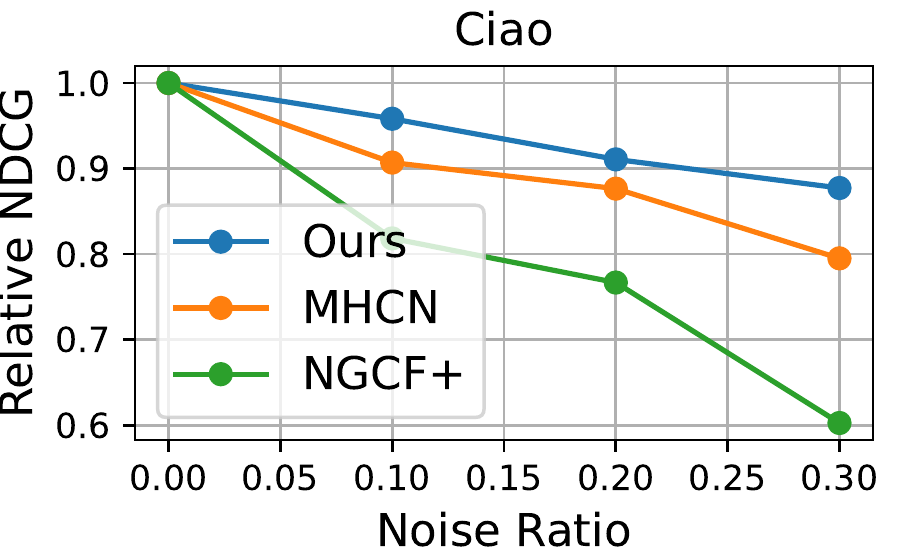}
    \includegraphics[width=0.19\textwidth]{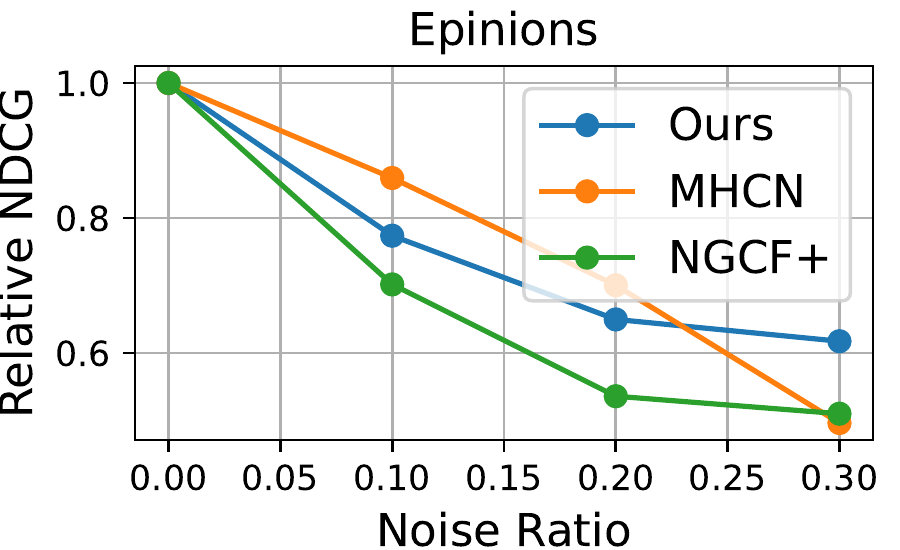}
    \includegraphics[width=0.19\textwidth]{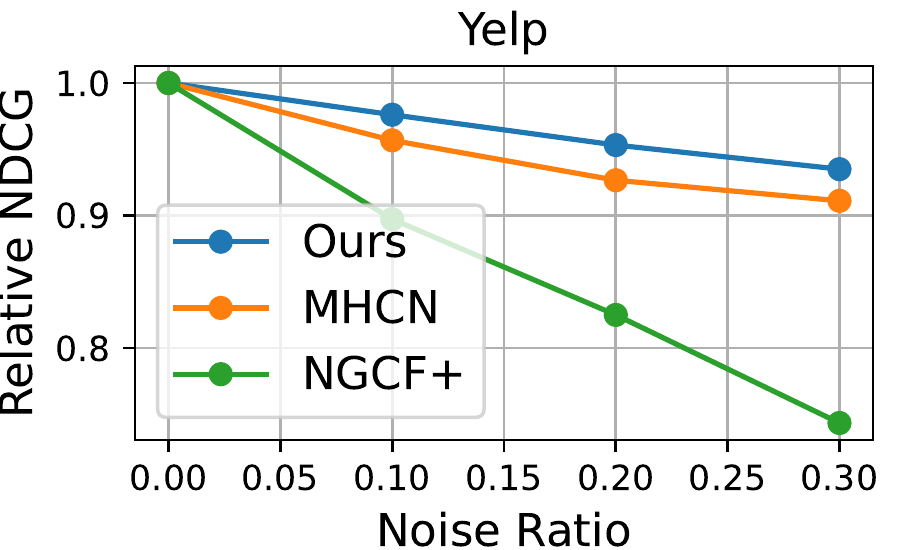}
    \includegraphics[width=0.19\textwidth]{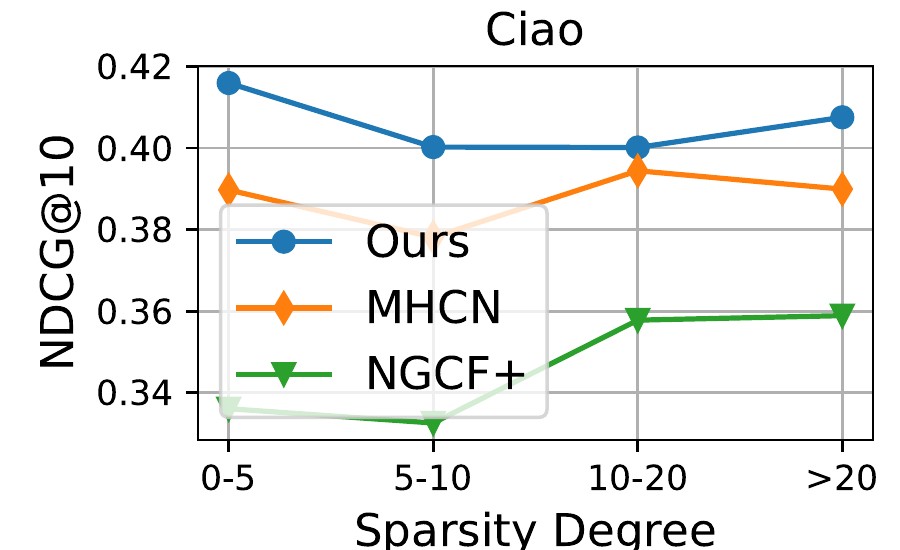}
    \includegraphics[width=0.19\textwidth]{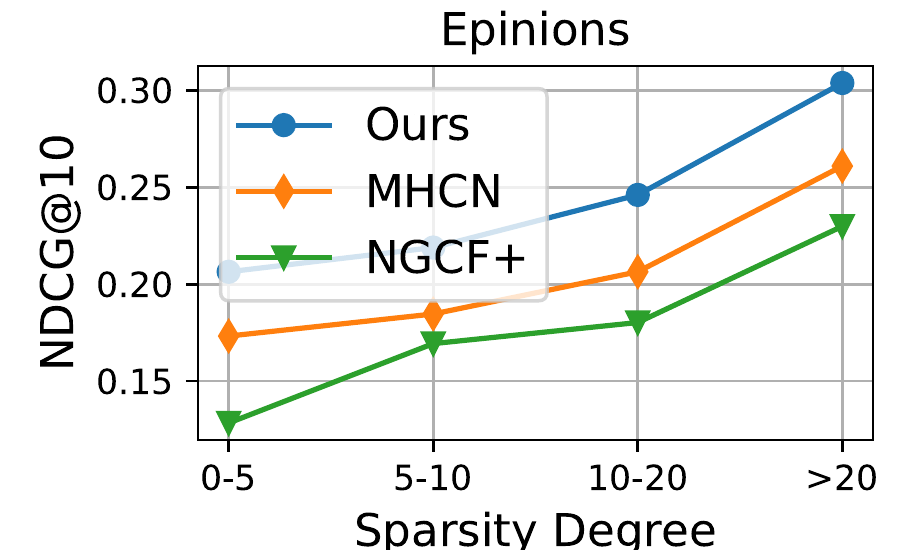}
\caption{Model robustness study \wrt~data noise and data sparsity, in terms of HR@N and NDCG@N.}
\label{fig:robustness}
\end{figure*}

\subsection{Model Robustness Evaluation (RQ3)}

In this section, we investigate the robustness of our \model\ against data sparsity and noise for recommendation. \\\vspace{-0.1in}

\paragraph{Data Sparsity.} To evaluate the model's performance on less active users with fewer item interactions, we partitioned the user set into four groups based on their node degrees in the user-item interaction graph $\mathcal{G}_{r}$, namely (0,5), [5,10), [10,15), and [20, $+\infty$). We separately measured the recommendation accuracy for each user group, and the evaluation results are reported in Figure~\ref{fig:robustness}. We observed that \model\ outperformed the best-performing baseline MHCN in most cases, further validating the effectiveness of our incorporated self-supervision signals for data augmentation under interaction label scarcity. \\\vspace{-0.1in}



\paragraph{Data Noise.} To investigate the influence of noisy effects on model performance, we randomly generated different percentages of fake edges (\ie, $10\%, 20\%, 30\%$) to create a corrupted interaction graph as noise perturbation. The relative performance degradation with different noise ratios is shown in Figure~\ref{fig:robustness}. Our \model\ demonstrates great potential in addressing data noise issues compared to competitors. We attribute this superiority to two reasons: 1) Graph structure learning with social relationships as self-supervision signals, which may alleviate the heavy reliance on interaction labels for representation learning. 2) The cross-domain self-augmented learning distills useful information from the interaction view to debias social connections and learn accurate social-aware collaborative relationships. \\\vspace{-0.1in}



\noindent These findings address the research question raised in RQ3 by demonstrating that our proposed \model\ improves the model's generalization ability over noisy and sparse data.

\begin{figure}[htbp]
\centering
\includegraphics[width=0.47\textwidth]{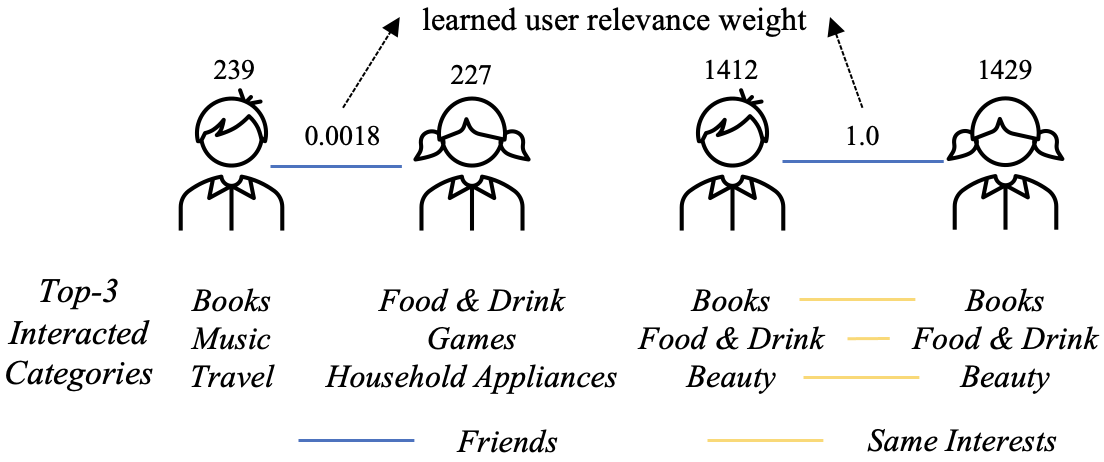}
\caption{Case study on denoising social relations for modeling user-item interaction patterns with sampled socially-connected user pairs.}
\label{fig:case}
\end{figure}

\subsection{Efficiency Analysis (RQ4)}
We conduct additional experiments to evaluate the efficiency of our method for model training when \model\ competes with baselines. We measure the computational costs (running time) of different methods on an NVIDIA GeForce RTX 3090 and present the training time for each model in Table~\ref{tab:efficiency}. The training cost of \model\ is significantly lower than most of the compared baselines, demonstrating its potential scalability in handling large-scale datasets in real-life recommendation scenarios. While existing social recommenders (\eg, MHCN and SMIN) leverage SSL for data augmentation, blindly maximizing the mutual information between user embeddings may lead to additional computational costs. In \model, we enhance the social-aware self-supervised learning paradigm with an adaptive denoising module. This simple yet effective framework not only improves the recommendation quality but also shows an advantage in training efficiency.


\begin{table}[t]
\small
\scalebox{0.9}
{\begin{tabular}{|c|c|c|c|c|c|c|}
\hline
Data     & DiffNet & NGCF+ & MHCN & KCGN  & SMIN & Our  \\ \hline
Ciao     & 8.1     & 8.2   & 4.92 & 26.9  & 7.8  & 3.2  \\ \hline
Epinions & 39.1    & 16.3  & 9.34 & 49.4  & 19.7 & 6.1  \\ \hline
Yelp     & 692.9   & 124.6 & 56.2 & 132.5 & 75.3 & 58.6 \\ \hline
\end{tabular}}
\caption{Model computational cost measured by running time (s).}
\label{tab:efficiency}
\end{table}

\subsection{Case Study}
In this section, we conduct a case study in Figure~\ref{fig:case} to qualitatively investigate the effects of our cross-view self-augmented learning framework in denoising social connections for user preference learning. Specifically, we sample two user-user pairs from the Ciao dataset and show the top-$3$ frequently interacted item categories for each user. From the figure, we observe that the social influence between user 239 and user 227 is identified as weak (\ie, learned lower user relevance weight) with respect to their interaction preference. This is manifested by the fact that they mainly interacted with different item categories, \ie, user 239: Books, Music, and Travel; user 227: Food \& Drink, Games, and Household Appliances. On the other hand, the social influence between user 1412 and user 1429 is learned to be strong (\ie, learned higher user relevance weight). Most of their interacted items come from the same categories, \ie, Books, Food \& Drink, and Beauty, indicating their similar preferences. This observation aligns with our expectation that \model\ can denoise social connections and encode social-aware user interests through meaningful SSL-enhanced representations for recommendation.


%% file: conclusion.tex
\section{Conclusion}
\label{sec:conclusion}

In this work, we propose a universal denoised self-augmented learning framework that not only incorporates social influence to help understand user preferences but also mitigates noisy effects by identifying social relation bias and denoising cross-view self-supervision. To bridge the gap between social and interaction semantic views, the framework introduces a learnable cross-view alignment to achieve adaptive self-supervised augmentation. Experimental results show that our new \model\ leads to significant improvements in recommendation accuracy and robustness compared to existing baselines. Additionally, the component-wise effects are evaluated with ablation study. In future work, we aim to investigate the incorporation of interpretable learning over diverse relations to improve the explainability of denoised self-supervised learners for recommendation. Such incorporation can provide insights into the decision-making process of the social-aware recommender system, enabling users to understand how the system arrives at its recommendation results.
